\documentclass[10pt,superscriptaddress,twocolumn,amsmath,amssymb,aps,pra]{revtex4}
\usepackage{mathrsfs}
\usepackage{graphicx}
\usepackage{dcolumn}
\usepackage{bm}

\newcommand{\be}{\begin{equation}}
\newcommand{\ee}{\end{equation}}
\newcommand{\bea}{\begin{eqnarray}}
\newcommand{\eea}{\end{eqnarray}}

\begin{document}
\title{The Dynamical Stripes in Spin-Orbit Coupled Bose-Einstein Condensates with Josephson Junctions}
\author{Chunyuan Shan}
\author{Xiaoyu Dai}
\author{Boyang Liu}\email{boyangleo@gmail.com}
\affiliation{Institute of Theoretical Physics, Faculty of Science, Beijing University of Technology, Beijing, 100124, China}

\date{\today}
\begin{abstract}
The Josephson dynamics of the Bose-Einstein condensation with Raman-induced spin-orbit coupling is investigated. A quasi-1D trap is divided into two reservoirs by an optical barrier. Before the tunneling between the reservoirs is turned on, the system stays in its equilibrium ground state. For different spin-orbit coupling parameters and interaction strengthes, the ground state displays a rich phase diagram. In this work we focus on the plane wave phase and the stripe phase. Our calculation shows that, when the tunneling is turned on, the plane wave phase evolves into a dynamical stripe phase, that is, the density of the particle changes from uniform to periodically modulated. Basically, this stripe is described by a sine function and the wave length, the amplitude and the initial phase of the function are all varying with time. If the system stays in stripe phase initially, the stripes become ``sliding" when the tunneling is turned on, which reflects the running of one of the phases of the wave function.
\end{abstract}
 \maketitle
\section{Introduction}
Tunneling is an intriguing and fundamental phenomenon in quantum mechanics.
A famous example is the Josephson effect \cite{Josephson1962,Josephson1974}, which has been widely studied in solid state
superconductors \cite{Giaever1960,Likharev1979}, superfluid helium \cite{Pereverzev1997,Hoskinson2005,Wheatley1975}. The Josephson effect manifests itself as an oscillation between two macroscopic wave functions. Such a macroscopic quantum phenomenon has applications in various fields, for instance, the quantum information and computation \cite{Martinis2002,Paik2011}, and precision measurements \cite{Makhlin2001}.
Recently, an analogue of the Josephson effect in ultracold atomic gases has been observed in both bosonic \cite{Albiez2005,Levy2007} and fermionic systems \cite{Valtolina2015,Burchianti2018}. Furthermore, experiments of cold atoms also explored the dynamics of superfluids in multiply connected traps with Josephson junctions. It is an important step in the construction of ultracold atomic ``circuits" and the the realization of atomtronic quantum interference device (AQUID) \cite{Ramanathan2011,Ryu2013,Ryu2020}.

On another front, based on the Berry phase effect the synthetic spin-orbit (SO) coupling has been generated in ultracold atoms using atom light interaction. For instance, the so-called ``Raman-induced SO coupling" has been realized in both bosonic \cite{Lin2011} and fermionic systems \cite{Wang2012,Cheuk2012}, which is a coupling between spin and motion of atoms in one spatial dimension. Recently, the two-dimensional SO coupling has also been realized in cold atoms \cite{Wu2016,Sun2018,Huang2016,Meng2016}. These progresses have stimulated intensive studies in this area \cite{Goldman2014,Zhai2015}. While the Josephson effect has been well studied in the conventional BEC and fermionic superfluids, the Josephson dynamics of SO coupled systems was also discussed in several theoretical works \cite{Zhang2012,March2014,Citro2015}.

In this work we investigate Josephson dynamics of the Bose-Einstein condensates (BECs) with Raman-induced SO coupling in an elongated trap, which is divided into two reservoirs by an optical barrier as sketched in Fig. \ref{fig:oscPRR} (a). The BEC with Raman-induced SO coupling has rich zero-temperature phase diagram, which has been determined in Refs.\cite{Ho2011,Li2012}. For certain parameters the energy spectrum of the system has two minima, the BEC can condense at one minimum to form a ``plane wave" phase, or it can condense at both of the minima to form a ``stripe" phase. Compared to the conventional BECs it's interesting to investigate the Josephson dynamics of BECs with more complex phase diagrams. Here we show how the system evolves when the tunneling between the two reservoirs is turned on. First, if the system stays in the plane wave phase initially, one observes that it will evolve into a dynamical stripe phase. The particle density changes from uniform to a periodically modulated pattern, a sine function. By ``dynamical" we mean that the amplitude, wave length and the initial phase of the sine function are all varying with respect to time. Second, if the system stays in the stripe phase initially, then it becomes ``sliding" when the tunneling is turned on. The variations of amplitude and wave length of the stripe are barely observable. However, the initial phase keep running with time, then the stripe exhibits a ``sliding" behavior. These phenomena are new state of matter, which have not been observed in experiments. They may stimulate potential applications in future.

Our work is organized as the followings. In Sec. \ref{sec:model} we introduce the Hamiltonian that describes the SO coupled BECs and the tunneling between them. Furthermore, to study the time evolution of the system, the mean-field Lagrangian and Eular-Lagrange equations are derived. In Sec. \ref{sec:plane} we show how the system varies from the plane wave phase when the tunneling is turned on. In Sec. \ref{sec:stripe} we show how the system varies from the stripe phase. Finally, Sec. \ref{sec:con} provides our conclusions. In Appendix \ref{sec:App} we present the explicit forms of the Eular-Lagrange equations of the system.

\section{Model\label{sec:model}}
In the realistic experiment the Josephson dynamics can be studied in a setup presented in Ref. \cite{Valtolina2015}, where a elongated trap is divided into two reservoirs using a thin optical barrier. We consider a model similar to this setup. Then the Hamiltonian can be cast into three parts ($\hbar=1$)
\bea
H=H_L+H_R+H_T,
\eea
where $H_j (j=L,R)$ describes SO coupled BEC in the $j$-th reservoir and is written as
\bea
&&H_j=\cr&&\int_{{\rm Re}_j} dx\Big\{\Psi^\dagger_j(x)h_0\Psi_j(x)+\frac{g_{\uparrow\uparrow}}{2}|\psi_{j\uparrow}(x)|^4
+\frac{g_{\downarrow\downarrow}}{2}|\psi_{j\downarrow}(x)|^4\cr&&+g_{\uparrow\downarrow}|\psi_{j\uparrow}(x)|^2|\psi_{j\downarrow}(x)|^2
\Big\}. \eea  The reservoirs are one dimensional and the integration above is taken in the region ${\rm Re}_j$ for the $j$-th reservior, where ${\rm Re}_L=[-l,0]$ and ${\rm Re}_R=[0,l]$ and $l$ is the length of the two reservoirs. Two-component Bose gases are prepared in the reservoirs with SO coupling generated via Raman coupling. $\psi_{j\uparrow}(x)$ and $\psi_{j\downarrow}(x)$ are the annihilation operators of the two components, and $\Psi_j(x)=\left(\begin{array}{cc}
\psi_{j\uparrow}(x) & \psi_{j\downarrow}(x)\end{array}\right)^T$. The interaction coupling constants between the bosons are denoted as $g_{\uparrow\uparrow}$, $g_{\downarrow\downarrow}$ and $g_{\uparrow\downarrow}$. In this work we consider the spin symmetric case, and hence we set $g_{\uparrow\uparrow}=g_{\downarrow\downarrow}=g$. The single-particle Hamiltonian $h_0$ with Raman-induced SO coupling is written as the following
\bea
h_0=\left(\begin{array}{cc} \frac{(-i\partial_x +k_0)^2}{2m} & \frac{\Omega}{2} \\ \frac{\Omega}{2} & \frac{(-i\partial_x -k_0)^2}{2m} \end{array}\right),
\eea where $\Omega$ is the strength of the Raman coupling and $k_0$ is the wave vector of the laser. The single particle dispersion has two branches, which can be derived as the following
\bea
E_{k,\pm}=\frac{k_x^2}{2m}\pm\sqrt{\frac{k_x^2k_0^2}{m^2}+\frac{\Omega^2}{4}}.
\eea For $\Omega<4E_r$, where $E_r=k_0^2/2m$ is the recoil energy, the lower branch $E_{k,-}$ displays two degenerate minima as sketched in Fig.\ref{fig:oscPRR} (a), Fig.\ref{fig:oscPLR} (a) or Fig.\ref{fig:oscST} (a).

The tunneling part $H_T$ describes the weak link between the two reservoirs, and is written as
\bea
H_T=K\Big(\Psi^\ast_L(0)\Psi_R(0)+\Psi^\ast_R(0)\Psi_L(0)\Big),
\eea where $K$ is the tunneling parameter and the tunneling of the bosons occurs at the location $x=0$.

When the tunneling between the two reservoirs is turned off, the system stays in the ground state. In this work we consider the situation that the lower branch of the single particle dispersion has double degenerate minima, then the condensate wave function can be written as the following,
\bea
\Psi_j(x)&&=\sqrt{n_j}e^{i\phi_j}\Bigg(\cos\alpha_j\left(\begin{array}{c}
\cos\theta_j \\ -\sin\theta_j\end{array}\right)e^{i(k_jx+\gamma_j)}\cr &&+\sin\alpha_j\left(\begin{array}{c}
\sin\theta_j \\ -\cos\theta_j\end{array}\right)e^{-i(k_jx+\gamma_j)}\Bigg), \label{eq:WF}
\eea
where $n_j$ is the number density of bosons, and $k_j$, $\alpha_j$ and $\theta_j$ are the variational parameters, which can be determined by minimizing the energy. For certain parameters the energy minimization generates $\alpha_j=0$ or $\pi/2$, then the condensate wave function is a plane wave with momentum $k_j$ or $-k_j$, which is the ``plane wave" phase \cite{Ho2011,Li2012}. For some parameters the energy minimization yields $\alpha_j=\pi/4$. It's a superposition of two wave function components with different momenta ($k_j$ and $-k_j$). Spatially it exhibits a periodic density modulation. This is the ``stripe phase". Furthermore, in order to study the Josephson effect, a overall phase $\phi_j$ and a relative phase $\gamma_j$ between the two parts with momentum $k_j$ and $-k_j$ have to be taken into account.

When the tunneling between the two reservoirs is turned on, the dynamics of the system can be derived from the action principle $\delta\int^{t_2}_{t_1} L dt=0$, where the Lagrangian is given by
\bea
L=\sum_{j=L,R}\int_{{\rm Re}_j} dx~ \Psi^\ast_ji\partial_t\Psi_j
-H. \eea
In the mean-field level the Lagrangian can be calculated by simply replacing the field $\Psi_j$ with the ground state wave function in Eq. (\ref{eq:WF}). Then it can be explicitly presented as the following,
\begin{widetext}
\bea
&&L =ln_0\Bigg\{z_p\dot{\delta \phi} -\left ( 1+z_p \right ) \left ( -(\dot{k_{L}}l/2-\dot{\gamma _{L} })\cos 2\alpha _{L}+\frac{k_{L}^2+k_{0}^2}{2m}+\frac{k_{L}k_{0}}{m}\cos 2\theta_{L}-\frac{\Omega }{2}\sin 2\theta_{L} \right )\cr&&-(1+z_p)^{2}\left( G_{1} \left( 1+\frac{1}{2}\sin^{2} 2\alpha _{L} \sin^{2} 2\theta _{L} \right )+G_{2} \left ( 1- \sin^{2} 2\alpha _{L} - \sin^{2} 2\theta _{L} +\sin^{2} 2\alpha _{L} \sin^{2} 2\theta _{L} \right )\right)\cr&&-\left ( 1-z_p \right ) \left ( (\dot{k_{R}}l/2+\dot{\gamma _{R} })\cos 2\alpha _{R}+ \frac{k_{R}^2+k_{0}^2}{2m}+\frac{k_{R}k_{0}}{m}\cos 2\theta_{R}-\frac{\Omega }{2}\sin 2\theta_{R} \right )\cr&&-(1-z_p)^2\left(G_{1} \left ( 1+\frac{1}{2}\sin^{2} 2\alpha _{R} \sin^{2} 2\theta _{R} \right )+G_{2} \left ( 1- \sin^{2} 2\alpha _{R} - \sin^{2} 2\theta _{R} +\sin^{2} 2\alpha _{R} \sin^{2} 2\theta _{R} \right )\right)\cr&&-2K^\prime \sqrt{1-z_p^2}  \bigg\{ \cos \delta \phi  \Big [ \cos \left ( \alpha_{R}-\alpha_{L}  \right ) \cos \left (  \theta_{R}-\theta_{L} \right )\cos \left (\gamma_{R}-\gamma_{L}\right )+\sin \left ( \alpha_{R}+\alpha_{L}  \right ) \sin \left (  \theta_{R}+\theta_{L} \right )\cos \left (\gamma_{R}+\gamma_{L}\right ) \Big ] \cr&&- \sin \delta \phi  \Big[ \cos \left ( \alpha_{R}+\alpha_{L}  \right ) \cos \left (  \theta_{R}-\theta_{L} \right )\sin \left (\gamma_{R}-\gamma_{L}\right )-\sin \left ( \alpha_{R}-\alpha_{L}  \right ) \sin \left (  \theta_{R}+\theta_{L} \right )\sin \left (\gamma_{R}+\gamma_{L}\right ) \Big]  \bigg\}\Bigg\},\label{eq:Lagrangian}
\eea
\end{widetext}
Here the total particle number is fixed as $(n_L+n_R)l=N$ and we define the average number density as $n_0\equiv N/2l$. Then, $n_L$ and $n_R$ are not independent, neither do $\phi_L$ and $\phi_R$. The Lagrangian only depend on the phase difference $\delta\phi=\phi_L-\phi_R$ and the particle number bias $z_p=(n_L-n_R)/2n_0$. The interaction parameters are defined as $G_1\equiv n_0(g+g_{\uparrow\downarrow})/4$ and $G_2\equiv n_0(g-g_{\uparrow\downarrow})/4$. The tunneling parameter is redefined as $K^\prime=K/l$.

In general we assume all the parameters $k_j$, $\alpha_j$, $\theta_j$, $\gamma_j$, $\delta\phi$, and  $z_p$ are time dependent and their temporal variation can be derived by the Euler-Lagrange equations
\bea
\frac{\partial L}{\partial \lambda_i}=\frac{\partial}{\partial t}\Big(\frac{\partial L}{\partial \dot{\lambda_i}}\Big),\label{eq:EL}
\eea
where $\lambda_i$ represent the parameters $k_j$, $\alpha_j$, $\theta_j$, $\gamma_j$, $\delta\phi$, $z_p$, and $\dot{\lambda_i}=\partial_t \lambda_i$. Please refer to the appendix A for the detailed equations.

\section{The Josephson Effect in the plane wave phase \label{sec:plane}}
\begin{figure}[h]
\includegraphics[width=0.48\textwidth]{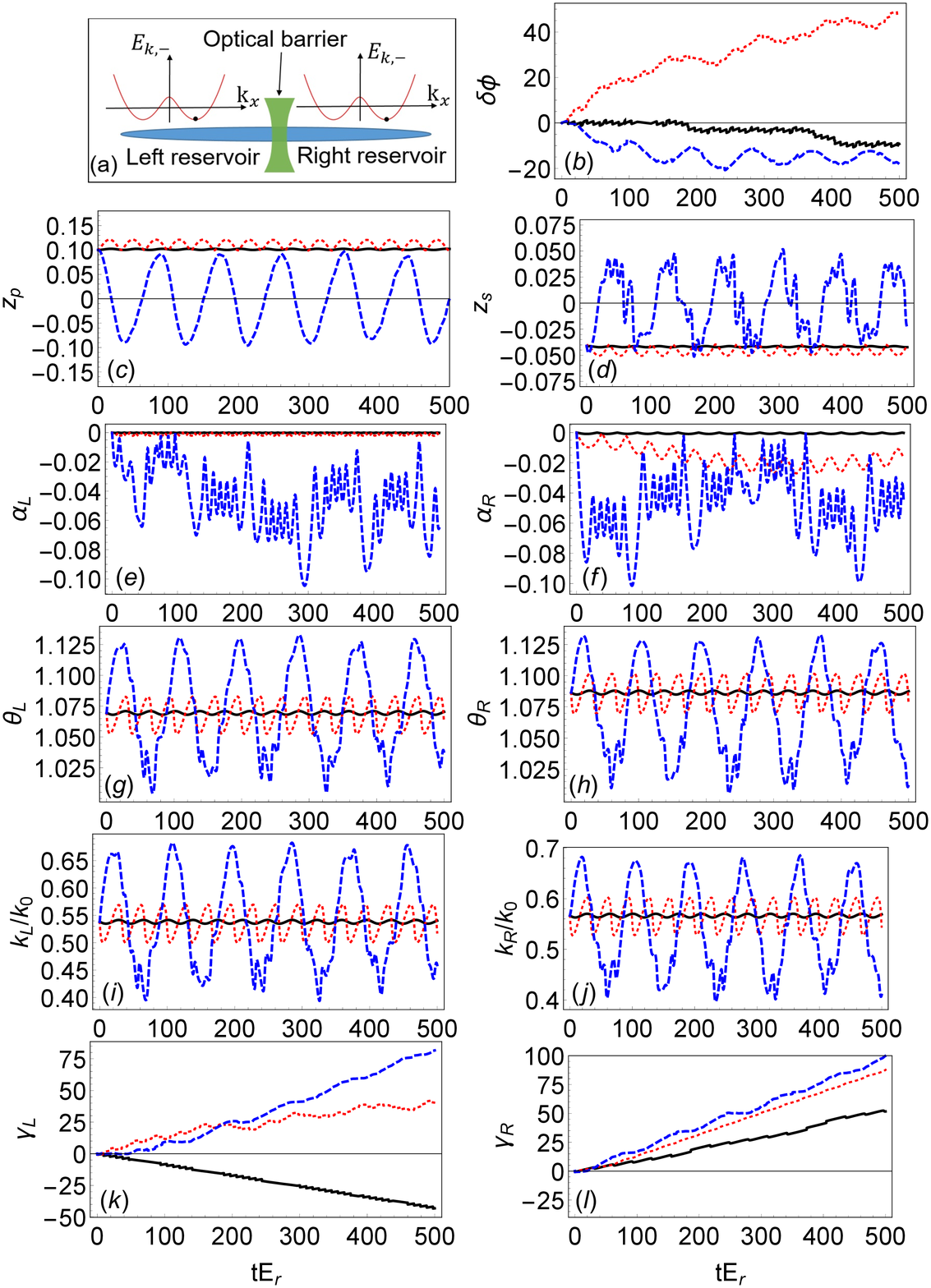}
\caption{(Color online) (a), the sketch of the initial state before the tunneling between the two reservoirs is turned on, where $k_L>0$ and $k_R>0$. (b)-(l), the temporal variations of the parameters. The black solid, red dotted and blue dashed curves are for $K^\prime=10^{-4}E_r, 10^{-3}E_r$, and $10^{-2}E_r$. $\Omega=3E_r$, $G_1=0.4E_r$ and $G_2=0.1E_r$ for all the graphs.}
\label{fig:oscPRR}
\end{figure}
\begin{figure}[h]
\includegraphics[width=0.48\textwidth]{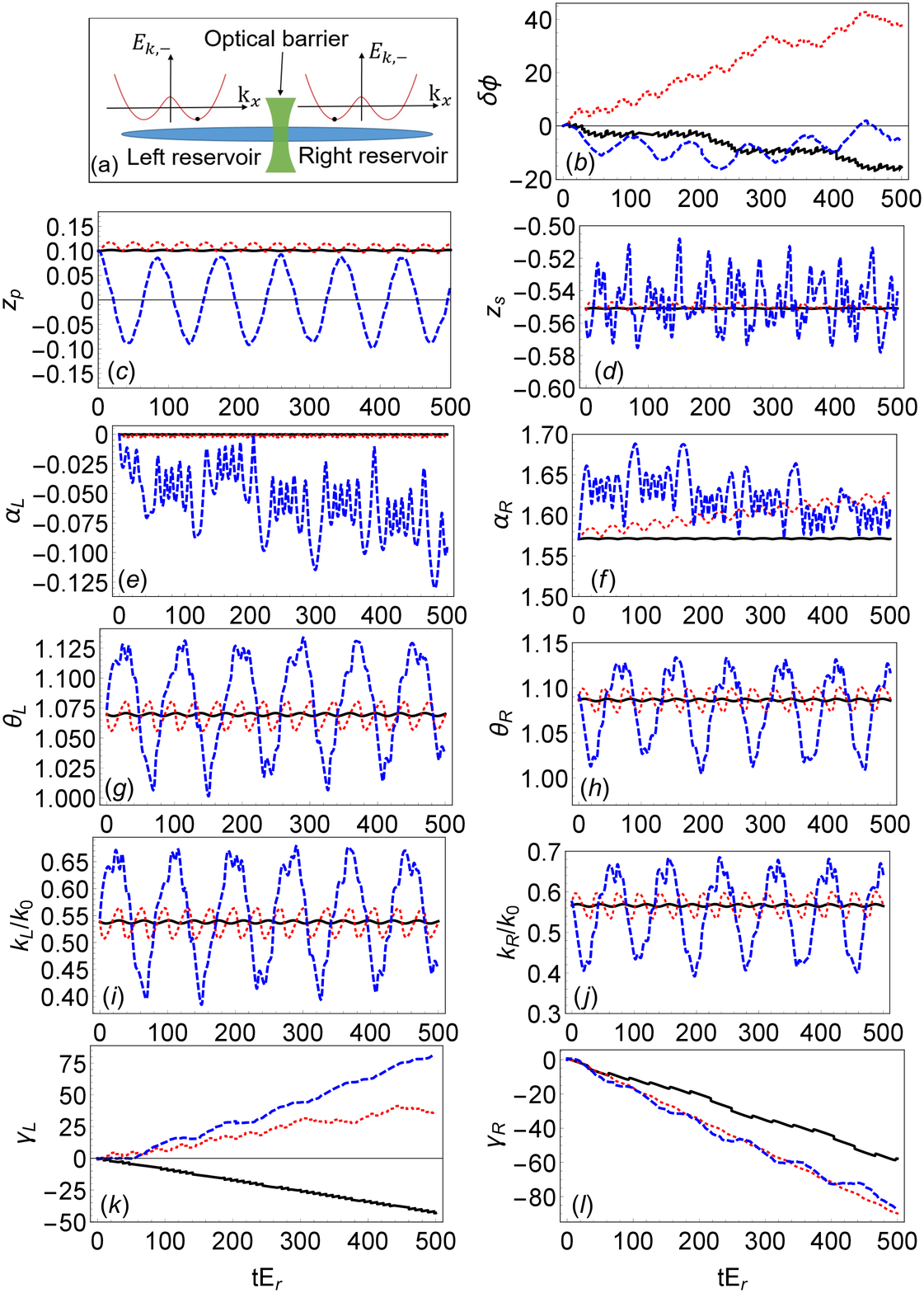}
\caption{(Color online) (a), the sketch of the initial state before the tunneling between the two reservoirs is turned on, where $k_L>0$ and $k_R<0$. (b)-(l), the temporal variations of the parameters. The black solid, red dotted and blue dashed curves are for $K^\prime=10^{-4}E_r, 10^{-3}E_r$, and $10^{-2}E_r$. $\Omega=3E_r$, $G_1=0.4E_r$ and $G_2=0.1E_r$ for all the graphs.  }
\label{fig:oscPLR}
\end{figure}

In this section we study the Josephson dynamics when the system stays in the plane wave phase initially. We will discuss two cases. First, the plane waves in both reservoirs are with momenta of the same sign, for instance, $k_L$ and $k_R$ are both positive as in Fig. \ref{fig:oscPRR} (a). Second, the momenta of the plane waves are of the opposite sign as shown in Fig. \ref{fig:oscPLR} (a). When the tunneling between the two reservoirs are turned on, the variations of all the parameters are demonstrated as in Fig. \ref{fig:oscPRR} and \ref{fig:oscPLR}. It's straight-forward to observe that, when the coupling strength $K^\prime$ is small, the parameters $k_j$, $\theta_j$ and $\alpha_j$ just oscillate around their values of the equilibrium state. The particle number bias $z_p$ exhibits the behavior of self-trapping for small $K^\prime$. As $K^\prime$ increases $z_p$ and $\delta\theta$ demonstrate the behavior of Josephson oscillation as shown in Fig. \ref{fig:oscPRR} (b), (c) and Fig. \ref{fig:oscPLR} (b), (c).
To study the spin behavior we define a spin bias as the following
\bea z_s &\equiv &\frac{(n_{L\uparrow}-n_{L\downarrow})l-(n_{R\uparrow}-n_{R\downarrow})l}{N}\cr
&=&\frac{1+z_p}{2}\cos2\theta_L\cos2\alpha_L-\frac{1-z_p}{2}\cos2\theta_R\cos2\alpha_R.\cr&&
\eea
The variations of $z_s$ are presented in Fig. \ref{fig:oscPRR} (d) and Fig. \ref{fig:oscPLR} (d). The major difference between the two cases is that $z_s$ oscillates around zero in Fig. \ref{fig:oscPRR} (d) and $z_s$ oscillates around a finite value in Fig. \ref{fig:oscPLR} (d). This difference reflects that the wave functions in the two reservoirs with the same momenta have the same spin configuration, while the ones with different momenta have different spin configuration.

\begin{figure}[h]
\includegraphics[width=0.48\textwidth]{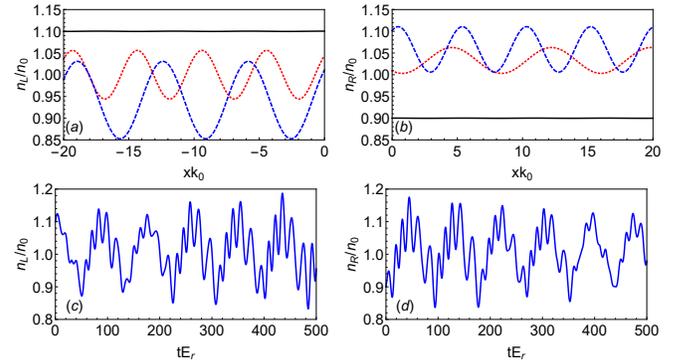}
\caption{(Color online) (a) and (b), the particle densities of left and right reservoirs, respectively, black solid, red dotted, and blue dashed curves are for $t=0$, $200/E_r$, and $400/E_r$. (c) and (d), the temporal variations of densities at locations $x=-10/k_0$ and $x=10/k_0$, respectively. $\Omega=3E_r$, $G_1=0.4E_r$, $G_2=0.1E_r$ and $K^\prime=10^{-2}E_r$ for all the graphs. }
\label{fig:nPLR}
\end{figure}

Next we study the temporal variation of the particle density for the case in Fig. \ref{fig:oscPLR} (a). There is no qualitative difference between the cases in Fig. \ref{fig:oscPRR} (a) and \ref{fig:oscPLR} (a), so we will not present the density plots of the case in Fig. \ref{fig:oscPRR} (a). The particle density of the two reservoirs can be expressed as the following,
\begin{align}
&n_{L/R}=\nonumber \\&n_0(1\pm z_p)(1+\sin2\alpha_{L/R}\sin2\theta_{L/R}\cos(2(k_{L/R}x+\gamma_{L/R})).\label{eq:density}
\end{align}
In Fig. \ref{fig:nPLR} (a) and (b) one observes that initially the density is uniform at $t=0$, which is the equilibrium-state plane wave. Then, when the tunneling between the two reservoirs is turned on, the density becomes periodically modulated. That is, the plane wave phase turns into a stripe phase. However, this stripe phase is different from the equilibrium-state stripe phase since it's dynamical. The shape of the stripe is still a sine function, however, the amplitude, the wave length and the initial phase of the sine function are varying with respect to time. If we study the density at particular locations, for instance, $x=-10/k_0$ and $10/k_0$, it is clearly illustrated that the oscillation is non-harmonic in Fig. \ref{fig:nPLR} (c) and (d).

Finally, we point out that the calculation will encounter a numerical breakdown for large tunneling parameter $K^\prime$. For instance, when $K^\prime$ increases to the order of $10^{-1}E_r$, some parameters quickly run into infinity. In this work we used an assumption that the wave function of the system is close to its equilibrium wave function. That is, the wave function in Eq. (\ref{eq:WF}) is in the same form as the one of the equilibrium state. The only thing different from the equilibrium state is that the parameters are all time dependent. The numerical breakdown implies our assumption here is only suitable for small $K^\prime$, the weak link case.

\section{The Josephson Effect in the stripe phase\label{sec:stripe}}
\begin{figure}[h]
\includegraphics[width=0.48\textwidth]{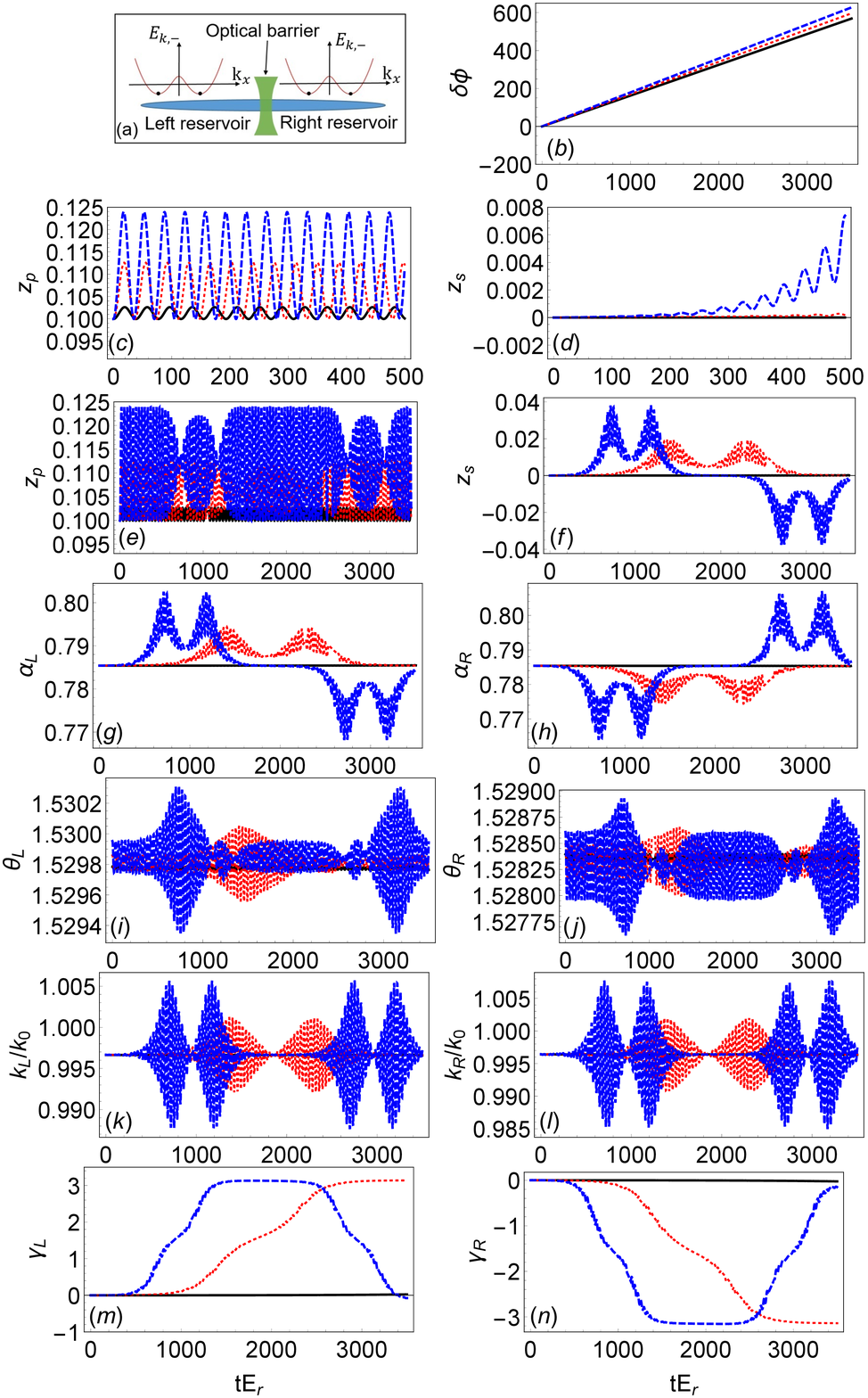}
\caption{(Color online) (a), the sketch of the initial state before the tunneling between the two reservoirs is turned on, both reservoirs are in the stripe phase. (b)-(n), the temporal variations of the parameters. The black solid, red dotted and blue dashed curves are for $K^\prime=10^{-4}E_r, 5\times10^{-4}E_r$, and $10^{-3}E_r$. $\Omega=0.4E_r$, $G_1=0.4E_r$ and $G_2=0.1E_r$ for all the graphs.  }
\label{fig:oscST}
\end{figure}
\begin{figure}[t]
\includegraphics[width=0.48\textwidth]{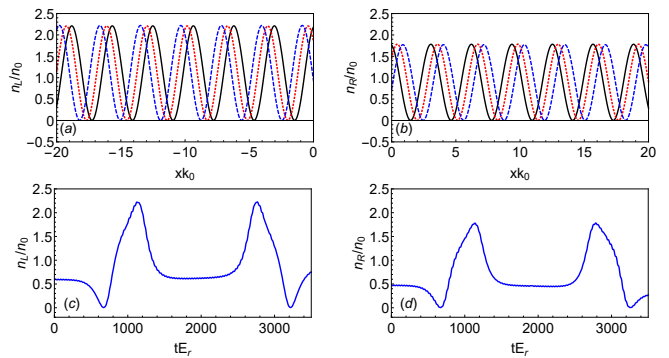}
\caption{(Color online) (a) and (b), the particle densities of left and right reservoirs, respectively, black solid, red dotted, and blue dashed curves are for $t=900/E_r$, $1100/E_r$, and $1200/E_r$. (c) and (d), the temporal variations of densities at location $x=-10/k_0$ and $x=10/k_0$, respectively. $\Omega=0.4E_r$, $G_1=0.4E_r$, $G_2=0.1E_r$ and $K^\prime=10^{-3}E_r$for all the graphs.  }
\label{fig:nST}
\end{figure}
Now we choose appropriate values for parameters $\Omega$, $k_0$, $G_1$ and $G_2$ so that the system stays in the stripe phase initially. When the tunneling between the reservoirs is turned on the variations of the parameters are demonstrated in Fig. \ref{fig:oscST}. Compared with the cases in Sec. \ref{sec:plane}, there are several differences. Firstly, the system is more fragile than it in Sec. \ref{sec:plane}. That is, the numerical breakdown occurs early than the cases in Sec. \ref{sec:plane} when the tunneling parameter $K^\prime$ increases. Hence, we only plot the cases for $K^\prime=10^{-4}E_r, 5\times10^{-4}E_r$ and $10^{-3}E_r$ in Fig. \ref{fig:oscST} since the numerical breakdown takes place in the order of $K^\prime=10^{-2}E_r$ or larger. Secondly, we plot $z_p$ and $z_s$ in Fig. \ref{fig:oscST} (c) and (d) for the same time scale as in Sec. \ref{sec:plane}. One observes that the particle number bias $z_p$ behaves similarly, while $z_s$ doesn't, which shows an oscillatory increasing behavior. Then we plot all the parameters in a larger time scale in Fig. \ref{fig:oscST} (e) to (n), where the parameters demonstrate a more complex pattern. Thirdly, the variations of the parameters $\alpha_j$, $\theta_j$ and $k_j$ are very small compared with the cases in Sec. \ref{sec:plane}. For instance, the ratio of the largest variation of $k_L$ with respect to its equilibrium-state value in Fig. \ref{fig:oscPRR} (i) is about $10\%$, while it's $0.8\%$ in Fig. \ref{fig:oscST} (k) when $K^\prime=10^{-3}E_r$.

In Fig. \ref{fig:nST} (a) and (b) we plot the particle densities of the two reservoirs. It is demonstrated that the shapes of the stripes are roughly the same for different time, that is, the stripes are sine functions with the same amplitudes and wave lengthes. Here we have to point out that they are not exactly the same. From Eq. (\ref{eq:density}) one can derive that the amplitude is $n_0(1\pm z_p)\sin2\alpha_{L/R}\sin2\theta_{L/R}$ and and the wave length is $\pi/k_{L/R}$. It has been shown that the variations of $z_p$, $\alpha_{L/R}$, $\theta_{L/R}$ and $k_{L/R}$ are all negligible. Hence, the changes of the amplitudes and the wave lengthes are barely observable. However, the phase $2\gamma_{L/R}$ is varying with respect to the time. Then, the stripes look like ``sliding". The phase $\gamma_{L/R}$ in Eq. (\ref{eq:WF})  actually corresponds to the translation of the wave functions. This is reason that the stripe slides when the phase $\gamma_{L/R}$ runs with time.

In Fig. \ref{fig:nST} (c) and (d)  the temporal variations of the densities of locations at $x=-10/k_0$ and $10/k_0$ are demonstrated, from which one  also observes the correspondence of the variations of densities and the running of phase  $\gamma_{L/R}$. For instance, when $400/E_r\lesssim t\lesssim 1200/E_r$ or $2600/E_r\lesssim t\lesssim 3200/E_r$ the phase $\gamma_{L/R}$ runs with time as shown in Fig. \ref{fig:oscST} (m) and (n), and the densities at locations $x=-10/k_0$ and $10/k_0$ also keep changing as illustrated in Fig. \ref{fig:nST} (c) and (d).  When $1600/E_r\lesssim t\lesssim 2400/E_r$,  $\gamma_{L/R}$ keeps a constant roughly, and the densities barely move either.

\section{Conclusions\label{sec:con}}
In summary, we study the Josephson dynamics of BECs with Raman-induced SO coupling in two reservoirs, which are connected by a weak link. In experiments, such a model can be realized in the setup similar to Ref.\cite{Valtolina2015}, where a cigar-shaped trap is divided into two reservoirs by an optical barrier. Initially, SO coupled BECs are prepared in the two reservoirs. We focus on two cases, first, the ground state of the system is in the plane wave phase, second, it's in the stripe phase before the tunneling between the two reservoirs is turned on. Then we investigate the time evolution of the system when the tunneling between the two reservoirs is turned on. Our results can be summarized as the followings. (i), starting from the plane wave phase, the system will evolve into a dynamical stripe phase. The particle density changes from uniform to periodically modulated. Basically, for any particular time the density is a sine function, however, when time changes, the wavelength, amplitude and initial phase all vary with time. (ii), if the system stays in the stripe phase initially, the stripe will slide when the tunneling is turned on. More precisely, in this case the wave length and the amplitude of the density wave are not observable and the initial phase keeps running. This running is reflected by the ``sliding" of the stripes.

\section{Acknowledgements}
The work is supported by the National Science Foundation of China (Grant No. NSFC-11874002), Beijing Natural Science Foundation (Grand No. Z180007).
\appendix
\section{Eular-Lagrange equations\label{sec:App}}
The mean-field Lagrangian is explicitly presented in Eq.(\ref{eq:Lagrangian}). Using Eq.(\ref{eq:EL}) all the Eular-Lagrange equations can be derived as the following,
\begin{widetext}
\bea
&&\delta \dot{\phi } +\frac{l}{2}\left (  \dot{k_{L}}\cos 2\alpha_{L} +\dot{k_{R}}\cos 2\alpha_{R}\right )+\dot{\gamma_{R}}\cos 2\alpha_{R}-\dot{\gamma_{L}}\cos 2\alpha_{L}\cr&&+\left ( \frac{k_{R}^2+k_{0}^2}{2m}+\frac{k_{R}k_{0}}{m}\cos 2\theta_{R}-\frac{\Omega }{2}\sin 2\theta_{R} -\frac{k_{L}^2+k_{0}^2}{2m}-\frac{k_{L}k_{0}}{m}\cos 2\theta_{L}+\frac{\Omega }{2}\sin 2\theta_{L}  \right )\cr&&+2G_{1} \bigg[ \left ( 1-z_p \right ) \left ( 1+\frac{1}{2}\sin ^2 2\theta_{R}\sin ^2 2\alpha _{R}  \right )-\left ( 1+z_p \right ) \left ( 1+\frac{1}{2}\sin ^2 2\theta_{L}\sin ^2 2\alpha _{L}  \right )   \bigg] \cr&&+2G_{2} \bigg[ \left ( 1-z_p \right ) \left ( 1-\sin ^2 2\theta_{R}-\sin ^2 2\alpha _{R}+\sin ^2 2\theta_{R}\sin ^2 2\alpha _{R}  \right )\cr&&-\left ( 1+z_p \right ) \left ( 1-\sin ^2 2\theta_{L}-\sin ^2 2\alpha _{L}+\sin ^2 2\theta_{L}\sin ^2 2\alpha _{L} \right )  \bigg] \cr&&+K^\prime \frac{2z_p}{\sqrt{1-z_p^2} } \bigg\{ \cos \delta \phi  \Big[ \cos \left ( \alpha_{R}-\alpha_{L}  \right ) \cos \left (  \theta_{R}-\theta_{L} \right )\cos \left (\gamma_{R}-\gamma_{L}\right )+\sin \left ( \alpha_{R}+\alpha_{L}  \right ) \sin \left (  \theta_{R}+\theta_{L} \right )\cos \left (\gamma_{R}+\gamma_{L}\right ) \Big]\cr&&- \sin \delta \phi  \Big[ \cos \left ( \alpha_{R}+\alpha_{L}  \right ) \cos \left (  \theta_{R}-\theta_{L} \right )\sin \left (\gamma_{R}-\gamma_{L}\right )-\sin \left ( \alpha_{R}-\alpha_{L}  \right ) \sin \left (  \theta_{R}+\theta_{L} \right )\sin \left (\gamma_{R}+\gamma_{L}\right ) \Big]  \bigg\}=0,
\eea
\bea
&&K^\prime\sqrt{1-z_p^2} \bigg\{ \sin \delta \phi  \Big[ \cos \left ( \alpha_{R}-\alpha_{L}  \right ) \cos \left (  \theta_{R}-\theta_{L} \right )\cos \left (\gamma_{R}-\gamma_{L}\right )+\sin \left ( \alpha_{R}+\alpha_{L}  \right ) \sin \left (  \theta_{R}+\theta_{L} \right )\cos \left (\gamma_{R}+\gamma_{L}\right ) \Big]\cr&&+ \cos \delta \phi  \Big[ \cos \left ( \alpha_{R}+\alpha_{L}  \right ) \cos \left (  \theta_{R}-\theta_{L} \right )\sin \left (\gamma_{R}-\gamma_{L}\right )-\sin \left ( \alpha_{R}-\alpha_{L}  \right ) \sin \left (  \theta_{R}+\theta_{L} \right )\sin \left (\gamma_{R}+\gamma_{L}\right ) \Big]  \bigg\}-\frac{\dot{z_p}}{2}=0,\cr&&
\eea
\bea
&&-\frac{l}{2}\dot{k_{L}}\left ( 1+z_p \right ) \sin 2\alpha_{L} +\left ( 1+z_p \right ) \dot{\gamma_{L} }\sin  2\alpha_{L} -\left ( 1+z_p \right )^{2}\left ( G_{1}+2G_{2}  \right )
 \sin^{2} 2\theta_{L}\sin 2\alpha_{L}\cos2\alpha_{L}\cr&&+ 2\left ( 1+z_p \right )^{2}G_{2} \sin 2\alpha_{L}\cos2\alpha_{L}-K^\prime \sqrt{1-z_p^{2} } \bigg\{ \cos \delta \phi  \Big[ \sin \left ( \alpha_{R}-\alpha_{L}  \right ) \cos \left (  \theta_{R}-\theta_{L} \right )\cos \left (\gamma_{R}-\gamma_{L}\right )\cr&&+\cos \left ( \alpha_{R}+\alpha_{L}  \right ) \sin \left (  \theta_{R}+\theta_{L} \right )\cos \left (\gamma_{R}+\gamma_{L}\right ) \Big]- \sin \delta \phi  \Big[ -\sin \left ( \alpha_{R}+\alpha_{L}  \right ) \cos \left (  \theta_{R}-\theta_{L} \right )\sin \left (\gamma_{R}-\gamma_{L}\right )\cr&&+\cos \left ( \alpha_{R}-\alpha_{L}  \right ) \sin \left (  \theta_{R}+\theta_{L} \right )\sin \left (\gamma_{R}+\gamma_{L}\right ) \Big]  \bigg\}
=0,
\eea
\bea
 &&\frac{l}{2}\dot{k_{R}}\left ( 1-z_p \right ) \sin 2\alpha_{R} +\left ( 1-z_p \right ) \dot{\gamma_{R} }\sin  2\alpha_{R}-\left ( 1-z_p \right )^{2}\left ( G_{1}+2G_{2}  \right )
 \sin^{2} 2\theta_{R}\sin 2\alpha_{R}\cos2\alpha_{R}\cr&&+ 2\left ( 1-z_p \right )^{2}G_{2} \sin 2\alpha_{R}\cos2\alpha_{R} -K^\prime \sqrt{1-z_p^{2} } \bigg\{ \cos \delta \phi  \Big[ -\sin \left ( \alpha_{R}-\alpha_{L}  \right ) \cos \left (  \theta_{R}-\theta_{L} \right )\cos \left (\gamma_{R}-\gamma_{L}\right )\cr&&+\cos \left ( \alpha_{R}+\alpha_{L}  \right ) \sin \left (  \theta_{R}+\theta_{L} \right )\cos \left (\gamma_{R}+\gamma_{L}\right ) \Big]+ \sin \delta \phi  \Big[ \sin \left ( \alpha_{R}+\alpha_{L}  \right ) \cos \left (  \theta_{R}-\theta_{L} \right )\sin \left (\gamma_{R}-\gamma_{L}\right )\cr&&+\cos \left ( \alpha_{R}-\alpha_{L}  \right ) \sin \left (  \theta_{R}+\theta_{L} \right )\sin \left (\gamma_{R}+\gamma_{L}\right ) \Big] \bigg\}
=0,
\eea
\bea
 &&\frac{\left ( 1+z_p \right ) }{2}\left ( \frac{2k_{L}k_{0}}{m}\sin 2\theta_{L} +\Omega \cos2\theta_{L}   \right )-\left ( 1+z_p \right )^{2}\left ( G_{1}+2G_{2}  \right )
 \sin^{2} 2\alpha_{L}\sin 2\theta_{L}\cos2\theta_{L}+ 2\left ( 1+z_p \right )^{2}G_{2}\sin 2\theta_{L}\cos2\theta_{L} \cr&&-K^\prime \sqrt{1-z_p^{2} } \bigg\{ \cos \delta \phi  \Big [ \cos \left ( \alpha_{R}-\alpha_{L}  \right ) \sin \left (  \theta_{R}-\theta_{L} \right )\cos \left (\gamma_{R}-\gamma_{L}\right )+\sin \left ( \alpha_{R}+\alpha_{L}  \right ) \cos \left (  \theta_{R}+\theta_{L} \right )\cos \left (\gamma_{R}+\gamma_{L}\right ) \Big ] \cr&&- \sin \delta \phi  \Big[ \cos \left ( \alpha_{R}+\alpha_{L}  \right ) \sin \left (  \theta_{R}-\theta_{L} \right )\sin \left (\gamma_{R}-\gamma_{L}\right )-\sin \left ( \alpha_{R}-\alpha_{L}  \right ) \cos \left (  \theta_{R}+\theta_{L} \right )\sin \left (\gamma_{R}+\gamma_{L}\right ) \Big]  \bigg\}
=0,
\eea
\bea
 &&\frac{\left ( 1-z_p \right ) }{2}\left ( \frac{2k_{R}k_{0}}{m}\sin 2\theta_{R} +\Omega \cos2\theta_{R}   \right )-\left ( 1-z_p \right )^{2}\left ( G_{1}+2G_{2}  \right )
 \sin^{2} 2\alpha_{R}\sin 2\theta_{R}\cos2\theta_{R}+ 2\left( 1-z_p \right )^{2}G_{2}\sin 2\theta_{R}\cos2\theta_{R} \cr&&-K^\prime \sqrt{1-z_p^{2} } \bigg\{ \cos \delta \phi  \Big[ -\cos \left ( \alpha_{R}-\alpha_{L}  \right ) \sin \left (  \theta_{R}-\theta_{L} \right )\cos \left (\gamma_{R}-\gamma_{L}\right )+\sin \left ( \alpha_{R}+\alpha_{L}  \right ) \cos \left (  \theta_{R}+\theta_{L} \right )\cos \left (\gamma_{R}+\gamma_{L}\right ) \Big] \cr&&- \sin \delta \phi  \Big[  -\cos \left ( \alpha_{R}+\alpha_{L}  \right ) \sin \left (  \theta_{R}-\theta_{L} \right )\sin \left (\gamma_{R}-\gamma_{L}\right )-\sin \left ( \alpha_{R}-\alpha_{L}  \right ) \cos \left (  \theta_{R}+\theta_{L} \right )\sin \left (\gamma_{R}+\gamma_{L}\right ) \Big] \bigg\}
=0,
\eea
\bea
\left ( 1+z_p \right ) \left ( \frac{k_{L}}{m}+\frac{k_{0}}{m}\cos 2\theta_{L} \right )
+\frac{1}{2}\dot{z_p}l \cos 2\alpha_{L}-l\left ( 1+z_p \right )\dot{\alpha_{L}}\sin 2\alpha_{L}=0,
\eea
\bea
\left ( 1-z_p \right ) \left ( \frac{k_{R}}{m}+\frac{k_{0}}{m}\cos 2\theta_{R} \right )
+\frac{1}{2}\dot{z_p}l \cos 2\alpha_{R}+l\left ( 1-z_p \right )\dot{\alpha_{R}}\sin 2\alpha_{R}=0,
\eea
\bea
&&\frac{\dot{z_p}}{2} \cos 2\alpha_{L}-\left ( 1+z_p \right ) \dot{\alpha_{L}}\sin 2\alpha_{L}-K^\prime \sqrt{1-z_p^2} \bigg\{ \cos \delta \phi  \Big[ \cos \left ( \alpha_{R}-\alpha_{L}  \right ) \cos \left (  \theta_{R}-\theta_{L} \right )\sin \left (\gamma_{R}-\gamma_{L}\right )\cr&&-\sin \left ( \alpha_{R}+\alpha_{L}  \right ) \sin \left (  \theta_{R}+\theta_{L} \right )\sin \left (\gamma_{R}+\gamma_{L}\right ) \Big]- \sin \delta \phi  \Big[ -\cos \left ( \alpha_{R}+\alpha_{L}  \right ) \cos \left (  \theta_{R}-\theta_{L} \right )\cos \left (\gamma_{R}-\gamma_{L}\right )\cr&&-\sin \left ( \alpha_{R}-\alpha_{L}  \right ) \sin \left (  \theta_{R}+\theta_{L} \right )\cos \left (\gamma_{R}+\gamma_{L}\right ) \Big]  \bigg\}=0,
\eea
\bea
&&-\frac{\dot{z_p}}{2} \cos 2\alpha_{R}-\left ( 1-z_p \right ) \dot{\alpha_{R}}\sin 2\alpha_{R}-K^\prime \sqrt{1-z_p^2} \bigg\{ \cos \delta \phi  \Big[ -\cos \left ( \alpha_{R}-\alpha_{L}  \right ) \cos \left (  \theta_{R}-\theta_{L} \right )\sin \left (\gamma_{R}-\gamma_{L}\right )\cr&&-\sin \left ( \alpha_{R}+\alpha_{L}  \right ) \sin \left (  \theta_{R}+\theta_{L} \right )\sin \left (\gamma_{R}+\gamma_{L}\right ) \Big]- \sin \delta \phi  \Big[ \cos \left ( \alpha_{R}+\alpha_{L}  \right ) \cos \left (  \theta_{R}-\theta_{L} \right )\cos \left (\gamma_{R}-\gamma_{L}\right )\cr&&-\sin \left ( \alpha_{R}-\alpha_{L}  \right ) \sin \left (  \theta_{R}+\theta_{L} \right )\cos \left (\gamma_{R}+\gamma_{L}\right ) \Big]  \bigg\}=0.
\eea
\end{widetext}

\end{document}